\documentstyle[12pt,aasms4,psfig]{article}

\def \cm3{cm$^{-3}$\thinspace }

\begin{document}

\title{ELECTRON-ION RECOMBINATION RATE COEFFICIENTS AND PHOTOIONIZATION
CROSS SECTIONS FOR ASTROPHYSICALLY ABUNDANT ELEMENTS VI. Ni II}
\author{Sultana N. Nahar$^1$ and Manuel A. Bautista$^2$}
\affil{$^1$ 
Department of Astronomy, The Ohio State University, Columbus, OH 43210}
\affil{$^2$
Centro de F\'{\i}sica, Instituto Venezolano de Investigaciones 
Cient\'{\i}ficas (IVIC),}
\affil{Altos de Pipe, Edo. Miranda, Vanazuela}                  

\begin{abstract}

We present the first detailed ab initio quantum mechanical calculations 
for total and state-specific recombination rate coefficients for 
e + Ni III $\rightarrow$ Ni II. These rates are obtained using a 
unified treatment for total electron-ion recombination that treats
the nonresonant radiative recombination and the resonant dielectronic 
recombination in a self-consistent unified manner in the close coupling
approximation. Large-scale calculations are carried out using a 
49-state wavefunction expansion from core configurations $3d^8$, 
$3d^74s$, and $3d^64p$ that permits the inclusion of prominent dipole 
allowed core transitions. These extensive calculations for the 
recombination rates of Ni II required hundreds  of CPU hours 
on the Cray T90. The total recombination rate coefficients are provided 
for a wide range of temperature. The state-specific recombination rates 
for 532 bound states of doublet and quartet symmetries, and the 
corresponding photoionization cross sections for leaving the core in 
the ground state, are presented. Present total recombination rate 
coefficients differ considerably from the currently used data in 
astrophysical models. 

\end{abstract}


\section{INTRODUCTION}

Nickel is one of the most important iron peak elements. For example,
radioactive decay Ni-Co-Fe is the primary power source for the light
curves of supernovae. Dipole allowed and forbidden Ni~II lines are often 
observed in the interstellar medium and H II regions, diffuse nebulae, 
novae, and supernova remnants. However, only a limited number of 
theoretical studies have been done for the atomic processes in Ni II.
With a Fe-like core, the ion poses a computational challenge with its 
strong electron-electron correlation effects. Bautista (1999) made the 
first detailed theoretical study for the radiative processes for 
bound-bound and bound-free transitions in Ni II presenting
total and partial photoionization cross sections for the ground and 
excited states.

In this report we study electon-ion recombination, e + Ni III 
$\rightarrow$ Ni II, and present the total recombination
rate coefficients, $\alpha_R(T)$. We use the unified treatment of 
total electron ion recombination of Nahar and Pradhan (1992, 1994, 1995).
The method involves large scale computations in the close coupling (CC)
approximation using the R-matrix method to (i)  provide 
recombination rate coefficients that are fully consistent 
with the photoionization cross sections of the ion, which allows, for
example, precise calculations of ionization balance, and (ii) treat 
nonresonant radiative and resonant dielectronic recombination in a 
self-consistent, unified manner, as explained in the first
paper of the series (Nahar \& Pradhan 1997). Present results are the
first detailed $\alpha_R(T)$ for recombination of Ni II. 

State specific recombination rate coefficients for 532 bound states of 
Ni II are presented. These are obtained from the {\it partial} 
photoionization cross sections including autoionizing resonances.
The partial cross sections correspond to photoionization leaving the
residual core in its ground state.

\section{THEORETICAL SUMMARY \protect\\}
\label{sec:level2}

The calculations are carried out in the close coupling approximation 
using the R-matrix method essentially as in the Opacity Project 
(Seaton 1987, Seaton et al. 1994). 
The theoretical details for photoionization cross sections for Ni II 
can be found in Bautista (1999).

The theory of unified treatment for total electron-ion recombination 
in the CC approximation is briefly outlined. The treatment considers 
the infinite number of states of the recombined ion and incorporates 
both the radiative recombination (RR) and the dielectronic recombination
(DR) processes in a self-consistent, unified manner. Details of 
the method are given in Nahar and Pradhan (1994, 1995) and in previous
papers in the present series. 

In the CC approximation the total wavefunction expansion, $\Psi(E)$, 
of the recombined ion, (N+1) electron system, is expanded in terms of 
the target (recombining ion) wavefunctions as:

\begin{equation}
\Psi(E) = A \sum_{i} \chi_{i}\theta_{i} + \sum_{j} c_{j} \Phi_{j},
\end{equation}

\noindent
where $\chi_{i}$ is the target wavefunction in a specific state
$S_iL_i\pi_i$ and $\theta_{i}$ is the wavefunction for the (N+1)th
electron in a channel labeled as $S_iL_i\pi_ik_{i}^{2}\ell_i(SL\pi)$; 
$k_{i}^{2}$ being its kinetic energy. $\Phi_j$ in the second summation 
are the bound channel functions of the (N+1)-electron system   
accounting for short range correlations and the orthogonality between 
continuum and bound orbitals. 

In the present approach the states of the recombined ion are divided 
into two groups, (A) low-n bound states, ranging from the ground  
to excited states with $n\leq n_{max}$ (typically  $n_{max}$ = 10), and 
(B) closely spaced high-n states, $n_{max}\leq n \leq \infty$. 

Group (A) bound states are treated via photo-recombination using detailed
balance with photoionization in the energy range where both the background
and resonant recombinations are important. The {\it partial} photoionization
cross sections leaving the core in the ground state are obtained for
all these bound states. The recombination cross sections, $\sigma_{RC}$, are
then obtained from the photoionization cross sections, $\sigma_{PI}$,
through principle of detailed balance:

\begin{equation}
\sigma_{RC} = {g_i\over g_f} {\hbar^2\omega^2\over{m^2v^2c^2}}\sigma_{PI}
\end{equation}

\noindent
where $g_i$ and $g_f$ are the statistical weight factors of the
recombining and recombined states respectively, $v$ is velocity of the 
photoelectron, and $\omega$ is the photon frequency. The recombination 
rate coefficient, $\alpha_R(T)$, at a given temperature is obtained by 
averaging $\sigma_{RC}(T)$ over the Maxwellian distribution of electrons, 
$f(v)$, as,

\begin{equation}
\alpha_{R}(T) = \int_0^{\infty}{vf(v)\sigma_{RC}dv}
= {g_i\over g_j}{2\over kT\sqrt{2\pi m^3kc^2T}}
\int^{\infty}_0E^2\sigma_{PI}(\epsilon) e^{-{\epsilon\over kT}}d\epsilon,
\end{equation}

\noindent
where $\epsilon$ is photoelectron energy, $E = \hbar \omega = 
\epsilon+I_p$ is the photon energy and $I_p$ is the ionization potential. 
As photoionization cross sections include the
detailed structures of autoionizing resonances, the sum of individual
rates of the bound states corresponds to inclusion of both the RR and
the DR in a unified manner.

The states in group (B) are dense and belong to small energy regions 
below target thresholds, $n_{max}<n\leq\infty$, where DR 
dominates. In this region, the outer electron interacts weakly with the 
core which decays through emission of a photon. As the photon energy  
approaches the excited core threshold, autoionization rate decreases as 
$\nu^{-3}$ (where $\nu$ is the effective quantum number of the Rydberg 
series belonging to the converging threshold) while the radiative decay 
process increases by the same order. These states are treated through 
precise DR theory by Bell \& Seaton (1985) as extended in Nahar \& Pradhan 
(1994) to obtain the DR collision strengths, $\Omega(DR)$. The
corresponding contributions to the total recombination rates are then 
obtained through Maxwellian average over $\Omega(DR)$.

\section{COMPUTATIONS\protect\\}
\label{sec:level2}

The first large-scale calculations are carried out for recombination of
Ni II in the close coupling approximation using the R-matrix method. 
The eigenfunction expansion of the Ni~II includes 49 states of core 
configurations $3d^8$, $3d^74s$, and $3d^64p$ (Bautista 1999). The 
expansion (listed in Table 1) allows for prominent dipole allowed core 
transitions, e.g. $3d^8(^3F) \rightarrow 3d^74p(^3G^o)$. The 
${4d}$ orbital is included in the correlation configurations $3d^7{4d}$ 
and $3p^53d^84d$. The target or the core wavefunctions were obtained 
using the atomic structure code SUPERSTRUCTURE (Eissner et al 1974). 
Observed energies are used for the target states for more accurate
resonance positions in photoionization cross sections. 
The second term in $\Psi(E)$ (Eq. 1) describing the short range 
electron correlation for Ni II configuration includes all possible 
configurations with maximum occupancies of $3d^{10}$, $4p^3$, and $4d^2$. 
The continuuam wavefunctions within the R-matrix boundary are represented 
by a basis set of 15 terms. Computations include partial waves with
$l_{max}$ = 10.

We obtained partial photoionization cross sections, $\sigma_{PI}$, 
leaving the core in the ground state for 532 low-$n$ bound states 
with $2S+1=$ 2 and 4, and $L \leq 7$, and $n\approx 10$, and $l \leq 
9$. Calculations for these cross sections are repeated 
for the present work at a finer energy mesh, and for values at high 
energy region. Calculations were carried out using the R-matrix 
package of codes developed by the Opacity Project (Seaton 1987, 
Berrington et al. 1987) that were extended for the Iron Project 
(Hummer et al 1993, Berrington et al. 1995). 

The recombination rate coefficients, Eq. (3), of individual bound states 
are obtained on averaging over $\sigma_{RC}(T)$ using the code RECOMB 
(Nahar 1996). The resonance structures in photoionization cross sections 
appear up to photon energies corresponding to the highest core state,
$3d^74p(^1D^o)$ (Table 1), through couplings of channels. At higher photon 
energies, the cross sections are extrapolated through fittings or by 
Kramers rule (Nahar \& Pradhan 1994). The sum of these individual rates
comprises the low-n contributions to the total recombiation rate 
coefficients, $\alpha_R(T)$. 

The collision strengths for dielectronic recombination, $\Omega(DR)$, 
for high-$n$ states are obtained using the same 49-state wavefunction 
as used for photoionization cross sections. We consider the channels 
radiatively decaying to core ground state, $3d^{8~3}F$, via dipole 
allowed transitions. There are 10 such core transitions (listed in
Table 2). The oscillator strengths for these transitions (in Table 2)
for $\Omega(DR)$ were obtained from SUPERSTRUCTURE (Eissner 
et al. 1974). The computations were carried out using the
extended code STGFDR (Nahar \& Pradhan 1994).

Independent close coupling calculations for ($e$+Ni~III) scattering,
using the same 49-state wavefunction, are 
also carried out for the electron impact excitation (EIE) collision
strength, $\Omega(EIE)$, at the target thresholds. These calculations
meet some consistency checks. 
It is important to determine the contributions of 
higher order multipole potentials since the present theory of 
DR collision strengths, based on multi-channel quantum defect theory,
neglects these contributions (Nahar \& Pradhan 1994). 
On the other hand, a proper choice of the R-matrix boundary and
size of R-matrix basis set can compensate for the difference introduced
by these potentials. We compute $\Omega(EIE)$ with and without the 
contributions from multipole potentials by a
switching parameter, ipert, in the program, STGF.
Values of $\Omega(EIE)$ excluding the contributions
(parameter ipert=0), and including them (ipert=1) are presented in Table 2. 
Good agreement between the two sets of values indicates 
that the validity of the DR calculations consistent with quantum defect
theory and the R-matrix calculations.  

Strong electron-electron correlations and channel couplings demanded 
extensive computational resources for this ion. The memory size and CPU 
time constraints on the Cray T90 forced us to compute cross sections for 
one symmetry at a time and in energy segments. For example, the time 
for computations of photoionization cross section for the $^2F$ states 
with a total of 127 channels, a Hamiltonian matrix of 1949$\times$ 
1949, and memory requirement of 45 MW, needed about 140 CPU hours. For 
overall computation the largest number of channels was 136 and the largest 
Hamiltonian matrix size was 2055 $\times$ 2055.  

On the other hand the calculation of DR collision strengths, 
$\Omega(DR)$, in the small energy region below the core thresholds, 
$\nu_{max}<\nu\leq\infty$, required less computational time ($\approx$ 40 
CPU hours in total) as no dipole matrix elements needed to be computed. 

The non-resonant background contributions from the high-n group 
(B) states to the total recombination is also included through a 
"top-up" scheme as explained in Nahar (1996). This contribution is 
significant at low temperatures.

\section{RESULTS AND DISCUSSIONS\protect\\}

The results of present calculations for Ni II and comparison with previous 
data are described in subsections below.

\subsection{Low-n States: Photoionization Cross Sections}

Partial photoionization cross sections ($\sigma_{PI}$) leaving the core
in the ground state are obtained for 532 bound states of Ni II. 
(The {\it total} cross sections are reported in Bautista (1999).) The 
cross sections exhibit extensive autoionizing resonances that enhance the 
overall photoionization rates considerably. 

State specific recombination rate coefficients are obtained from the
partial photoionization cross sections. The percentage contributions 
and relative weight of individual bound states to the total $\alpha_R$ 
varies with temperature due to resonance structures and enhancement 
in the background cross sections at different energies. Some 
important features of photoionization cross sections related to
recombination are illustrated in Figs. 1 and 2. 

Fig. 1 presents $\sigma_{PI}$ of five states. The ground state of 
Ni~II is an equivalent electron state, $3d^9(^2D)$, and is one of the 
dominant contributors at all temperatures because of its relatively 
large background photoionization cross sections even at high 
photon energies, as seen in the top panel, Fig. 1a. The state also
exhibits high resonances at relatively higher eneries. The first 
excited state, $3d^84s(^4F)$, is also a dominant state which shows 
extensive resonances and enhanced background in the high energy 
region (Fig. 1b). Cross sections of excited states, $3d^74s4p(^2G^o)$
and $3d^74s4p(^4G^o)$ in Figs.1 c and d are examples of states 
that dominate the low temperature region because of
their near threshold background raised by resonances, decay decay
at higher energies. Fig. 1e illustrates the dominance of the state
$3d^8(^3F)4d(^2G)$ at higher temperature by its dense and wide 
resonances in relatively high energy region. It is found that the 
summed contribution of the quartets dominates the total $\alpha_R(T)$
at lower temperature while that of the doublets contributes more at 
higher temperature. 

The existence of photo-excitation-of-core (PEC) resonances (Yu \&
Seaton 1987) at high energies is also a reason for increased total 
recombination rate coefficients at high temperatures. These relatively
wide resonances are observed in the photoionization cross sections 
of excited states with a valence electron. The PEC resonances are 
introduced when the core is excited via dipole allowed transitions, 
while the outer electron remains a spectator. Fig. 2 illustrates the wide
and prominent PEC resonances in the cross sections of Rydberg series
of states, $3d^8nd(^2F)$ where $5d \leq nd \leq 11d$, of Ni II. The 
arrows point the PEC positions at the core thresholds, such as, $^3G^o$, 
$^3D^o$, $^3F^o$, etc (listed in Table 2) for dipole allowed transitions 
of the core ground state, $^3F$. PEC resonances become more distinct with
higher $n$.

\subsection{High-n States: Collision Strengths for Dielectronic Recombination}

The DR collision strengths, $\Omega(DR)$, are obtained for recombination 
into the high-n states in the energy region $\nu_{max}<\nu\leq\infty$ 
below each state of the core ion contributing to DR. The resonances in
$\Omega(DR)$ correspond to Rydberg series of states $S_tL_t\pi_t\nu l$ 
converging to the threshold $S_tL_t\pi_t$. $\Omega(DR)$ is obtained in
two forms: with detailed energy variation with resonances, and analytically  
averaged over resonances. The features of $\Omega(DR)$ 
are illustrated in Fig. 3 where the dotted curves are the detailed 
ones with resonances, and the solid curves are resonance averaged. 
The lower panel presents $\Omega(DR)$ below the 10 core states,
thresholds for allowed transitions, while the upper 
panel presents the expanded features for the first three thresholds. 
The excited target thresholds, given in Table 2, are marked by arrows in
the figure.

As $\nu$ increases, the resonances appear closer while the background 
rises (as seen in the solid curve) such that $\Omega(DR)$ 
peaks sharply converging on to the threshold. That is, as $\nu$ increases 
the outer electron
remains loosely bound in a highly excited state while the core decays 
radiatively. $\Omega(DR)$ goes to zero beyond the threshold as the 
trapped electron flux in the closed channels below the threshold 
is released through excitation of the target state. We also note that
$\Omega(DR)$ is almost zero at the starting energy where $\nu$ = 10.0, 
indicating negligible radiation damping below this energy. 
For the total recombination rate coefficients presented here, we choose 
contrubutions from the resonance averaged $\Omega(DR)$, rather than 
the detailed one, because of higher numerical accuracy. 

In Fig. 3, the filled circles are the $\Omega(EIE)$, collision strengths 
for electron impact excitation (EIE), at the excited target thresholds, 
These are obtained from (e+ Ni III) scattering calculations in the close 
coupling approximaton, as mentioned before. The comparison of 
$\Omega(EIE)$ with 
$<\Omega(DR)>$ provide a check of conservation of total electron-photon 
flux such that the trapped electron flux due to DR resonances below a 
threshold equals that released due to EIE at the threshold, i.e., 
$<\Omega(DR)>$ should equal $\Omega(EIE)$ (without the multipole 
potentials ipert=0). Comparison between the two values given in Table 2 show
good agreement in general except at threshold $x^3G^o$ where the difference
is large. The discrepancy can be explained by the presence of a near 
threshold resonance in $\Omega(EIE)$ and insufficient energy resolution 
for both collision strengths.

\subsection{State-specific and total recombination rate coefficients}

We present both the state-specific and the total recombination rate 
coefficients for the recombined ion, Ni~II. The
total rates are obtained from the sum of the state specific rates 
along with the contributions from the high n states. The
individual state recombination rates of the bound states are of
importance in the determination of non-local thermodynamic equilibrium
(NLTE) level population and recombination line intensities. State 
specific rates are obtained for 532 bound states of Ni II with n 
$\leq$ 10 and $l \leq 9$. 

Table 3 presents the 20 dominant states in order of their contributions, 
for doublets and quartets separately, at temperatures T = 100, 1000, 
10000, and 20000 K. The order of the contributing states varies
with temperature depending on the resonance structures in $\sigma_{PI}$ 
with energies. The state specific rates at higher temperatures are 
somewhat underestimated since high-$n$ ($n > 10$) DR contributions are not 
included individually. It should be noted that state specific rate 
coefficients are different from  effective recombination rate coefficients, 
used in spectroscopic modeling, since the latter include contributions 
from higher bound states through radiative cascades.

The complete table for the state specific rates are available 
electronically. However, there may be some spectroscopic 
misidentification of states. The strong electron-electron correlation 
in Ni II introduces some complications in the quatum defect analysis for 
LS term identification. There are some cases when the same state can 
correspond to a few possible spectroscopic identifications, rather than 
a unique one. We also noted a few states missing in the calculated set
but have been observed.

Table 4 presents the total electron-ion recombination rate coefficients, 
$\alpha_R(T)$, of Ni~II recombining from Ni~III ground state.
The rates are given at temperatures from 10 to $10^7$K,
with a temperature mesh of $\Delta log_{10}T$ = 0.1. The solid curve
in Fig. 4 shows $log_{10} \alpha_R$ vs. $log_{10} T$ and the 
basic features. As for other ions, the total (e+ion) recombination rate 
for Ni II is maximum at low
temperatures due to dominance by radiative recombination. $\alpha_R(T)$ 
decreases by more than an order of magnitude as the temperature increases 
from $log_{10} T(K)$=1 to $log_{10}T(K) \approx$ 4.4, but rises again 
at higher temperature because of dominant contributions of DR which
peaks at about $10^5$ K, above which it falls smoothly. There is a 
small "bump" in the rates, the low-temperature DR bump, at around $10^3$K 
due to near threshold autoionizing resonances in the 
photoionization cross sections. This bump has been observed in several
other ions (e.g. Nahar \& Pradhan 1997). We note that the contributions 
from the nonresonant background cross sections of high-n states,
$n_{max} <n \leq \infty$, to the total $\alpha_R(T)$ is considerable at 
low temperatures when the mean electron energy is in general low for 
forming autoionizing resonances contributing to DR.

Present total $\alpha_R(T)$ (solid curve) are compared with the 
recombination rate coefficients derived by Shull and Steenberg 
(1982) in Fig. 4. They derived the radiative recombination rates (dashed 
curve) by extrapolation along isoelectronic sequences, and obtained
the dielectronic recombination rates (dotted curve) using the Burgess 
General formula (Burgess 1965). Their RR and DR rates have been added 
for the total (dot-dash curve). These values underestimate Ni II 
recombination rates in the low temperature region since the near
threshold resonance structures were not considered in their data
sources. The high temperature rates on the other hand are overestimated 
because the Burgess formula does not 
include: (i) the interference effect of the resonant DR and continuum
background which is considerable for multi-electron systems, and (ii) the 
autoionization into excited core states in addition to the ground state. 
This latter process reduces DR considerably if the channel couplings are 
strong, as in the present case.

The accuracy of the present total recombination rate coefficients, 
$\alpha_R(T)$, for Ni~II may be estimated to be about 30\% in most of the 
temperature range, except at very high temperatures where the cross
sections are extrapolated and at very low temperatures where resonances
may require higher resolution. 
The accuracy estimate is based on the general agreement 
between the calculated and the observed energies, use of observed target 
energies for accurate resonance positions, the general accuracy of 
the CC method for photoionization cross sections, electron scattering, 
and DR collision strengths, and previous comparison between results of 
the unified method and experiment (e.g. Zhang, Nahar \& Pradhan 1999). 

Some effects not included here could improve the results. First, the
relativistic effects could be important for this heavy, many electron ion. 
Although the unified treatment for (e+ion) recombination has been 
extended to include relativistic effects in the Breit-Pauli R-matrix
approximation (Zhang \& Pradhan 1997, Zhang, Nahar \& Pradhan 1999)
and employed for highly charged Li-like and He-like C and Fe (Nahar, 
Pradhan \& Zhang 2000, 2001) we do not expect relativistic effects to 
be too important
for Ni II. Also, by using  a larger wavefunction expansion that 
includes states for core transitions at higher energies may somewhat
enhance the recombination rates. However, additional calculations 
including these effects will require extensive computing resourses.

\section{CONCLUSION\protect\\}

Total and state specific electron-ion recombination rate coefficients, 
$\alpha_R(T)$, are presented for e + Ni III $\rightarrow$ Ni II. 
The calculations were carried out in the close coupling approximation 
employing a unified treatment. To our knowledge, this is the first 
detailed and accurate study for the recombination of Ni~II. The
partial photoionization cross sections of a large number of bound states
are also presented. Present recombination rates, alongwith
the {\it total} photoionization cross sections obtained with the same
close coupling expansion (Bautista 1999), will provide self-consistent 
atomic data for accurate calculations of ionization balance for plasmas in
photoionization or coronal equilibria. The total recombination rate
coefficients  differ considerably from the currently used values.

All data may be obtained from the first author by e-mail at: 
nahar@astronomy.ohio-state.edu.

\acknowledgments

This work was supported partially by the NSF (AST-987-0089) and the
NASA Astrophysics Data Program. The computations were carried out at 
the Center for Computational Science at NASA Goddard Space Flight 
Center, and at the Ohio Supercomputer Center in Columbus, Ohio. 


\begin{table}
\caption{The 49 LS terms and energies, $E_t$ (in Rydberg), of Ni III in the 
eigenfunction expansion of Ni II. 
\label{table1}}
\begin{tabular}{llcllcllc}
\tableline
\multicolumn{2}{c}{Term} & $E_t$ & \multicolumn{2}{c}{Term} & $E_t$ & 
\multicolumn{2}{c}{Term} & $E_t$ \\
\tableline
$3d^8$ & $^3F^e$ & 0.0 & 
$3d^74s$ & $^3F^e$ & 0.884557 &
$3d^74p$ & $^3G^o$ & 1.21295 \\
$3d^8$ & $^1D^e$ & 0.118808 & 
$3d^74s$ & $^1F^e$ & 0.920024 &
$3d^74p$ & $^1H^o$ & 1.21402  \\ 
$3d^8$ & $^3P^e$ & 0.144311 &
$3d^74p$ & $^5F^o$ & 1.00141 &
$3d^74p$ & $^1F^o$ & 1.22136 \\
$3d^8$ & $^1G^e$ & 0.201524 &
$3d^74p$ & $^5D^o$ & 1.01975 &
$3d^74p$ & $^3P^o$ & 1.22660  \\
$3d^8$ & $^1S^e$ & 0.469649 &
$3d^74p$ & $^5G^o$ & 1.02468 & 
$3d^74p$ & $^3D^o$ & 1.23933 \\
$3d^74s$ & $^5F^e$ & 0.490628 & 
$3d^74p$ & $^3G^o$ & 1.05115 &
$3d^74p$ & $^3G^o$ & 1.24696  \\
$3d^74s$ & $^3F^e$ & 0.558378 & 
$3d^74p$ & $^3F^o$ & 1.05715  & 
$3d^74p$ & $^3I^e$ & 1.24677 \\
$3d^74s$ & $^5P^e$ & 0.640931 &
$3d^74p$ & $^3D^o$ & 1.07862 &
$3d^74p$ & $^1S^o$ & 1.24983  \\
$3d^74s$ & $^3G^e$ & 0.679739 &
$3d^74p$ & $^5S^o$ & 1.10526  &
$3d^74p$ & $^3D^o$ & 1.25616 \\
$3d^74s$ & $^3P^e$ & 0.705399 & 
$3d^74p$ & $^5D^o$ & 1.13398 &
$3d^74p$ & $^1I^e$ & 1.26338  \\
$3d^74s$ & $^3P^e$ & 0.715485 &
$3d^74p$ & $^3S^o$ & 1.18346  &
$3d^74p$ & $^3F^o$ & 1.27101 \\
$3d^74s$ & $^1G^e$ & 0.713123 &
$3d^74p$ & $^3H^o$ & 1.19329 &
$3d^74p$ & $^3S^o$ & 1.27478  \\
$3d^74s$ & $^3H^e$ & 0.739700 &
$3d^74p$ & $^3F^o$ & 1.20136  &
$3d^74p$ & $^1P^o$ & 1.27960 \\
$3d^74s$ & $^3D^e$ & 0.742560 &
$3d^74p$ & $^5P^o$ & 1.20340 &
$3d^74p$ & $^3H^o$ & 1.28986 \\
$3d^74s$ & $^1P^e$ & 0.761912 &
$3d^74p$ & $^3P^o$ & 1.20688  &
$3d^74p$ & $^1D^o$ & 1.28890 \\
$3d^74s$ & $^1H^e$ & 0.773121 &
$3d^74p$ & $^1G^o$ & 1.20589 &
 & & \\
$3d^74s$ & $^1D^e$ & 0.780518 &
$3d^74p$ & $^3D^o$ & 1.20764  &
 & & \\
\tableline
\end{tabular}
\end{table}

\begin{table}
\caption{The transition probabilities ($A_{ji}$) from the excited 
thresholds to the ground $a^3F$ state of target Ni III. Next three column 
list the peak values of DR collision strengths, $<\Omega(DR)>$, and
electron-impact excitation collision strengths, $\Omega(EIE)_0$ (0-excluding 
contributions of multipole potentials) and $\Omega(EIE)_1$ (1-including these 
contributions), at the these thresholds.
\label{table2}}
\begin{tabular}{lcccc}
\tableline
Transition & $A_{ji}$(a.u.) & $<\Omega (DR)>$ & $\Omega (EIE)_0$ & 
$\Omega(EIE)_1$ \\
\tableline
$a^3F^e \leftarrow z^3G^o$  & 1.98(-9) & 2.25   & 3.76 & 3.76 \\
$a^3F^e \leftarrow z^3F^o$  & 1.86(-8) & 6.23 & 6.52 & 6.52 \\
$a^3F^e \leftarrow z^3D^o$  & 1.62(-8) & 3.62 & 3.45& 3.68\\
$a^3F^e \leftarrow y^3F^o$  & 2.27(-8) & 4.60 & 4.54 & 4.54 \\
$a^3F^e \leftarrow y^3D^o$  & 4.84(-9) & 0.957& 0.941& 0.941\\
$a^3F^e \leftarrow y^3G^o$  & 3.93(-9) & 1.24  & 1.21 & 1.21 \\
$a^3F^e \leftarrow x^3D^o$  & 4.49(-9) & 0.80  &0.793& 0.793\\
$a^3F^e \leftarrow x^3G^o$  & 2.42(-8) & 0.678 & 2.10 & 2.10 \\
$a^3F^e \leftarrow w^3D^o$  & 2.82(-8) & 3.05 & 3.09 & 3.09 \\
$a^3F^e \leftarrow v^3D^o$  & 4.53(-9) & 0.981& 0.791& 0.791\\
\tableline
\end{tabular}
\end{table}


\begin{table}
\caption{Individual state recombination rate coefficients (in units of 
$cm^3s^{-1}$) for e + Ni III $\rightarrow$ Ni II at temperatures, T = 100, 
1000, 10000, and 20000 K. The first 20 dominant doublets and quartets
are listed in order of their contributions. Their sum and percentage to
the total are specified below. Notation a-b means a$\times 10^{-b}$. }
\scriptsize
\begin{tabular}{lcclcclcclcc}
\tableline
\multicolumn{3}{c}{100\,K} & \multicolumn{3}{c}{1000\,K}& 
\multicolumn{3}{c}{10000\,K} & \multicolumn{3}{c}{20000\,K} \\
\multicolumn{2}{c}{State} & $\alpha_R$ & \multicolumn{2}{c}{State} & 
$\alpha_R$ & \multicolumn{2}{c}{State} & $\alpha_R$ & 
\multicolumn{2}{c}{State} & $\alpha_R$ \\
\tableline
\multicolumn{12}{c}{Doublets} \\
 $3d^74s^2       $ & $^2H^e$ & 4.99-13& $3d^8~  ^3P^e 4s$ & $^2P^e$ & 6.69-13& $
3d^74s ^3F^e 4p$ & $^2G^o$ & 2.07-13& $3d^9           $ & $^2D^e$ & 4.82-13 \\
 $3d^8~  ^1G^e 4p$ & $^2F^o$ & 2.99-13& $3d^74s^2       $ & $^2P^e$ & 4.39-13& $
3d^74s ^3F^e 4p$ & $^2D^o$ & 1.06-13& $3d^8  ~^3F^e 4d$ & $^2G^e$ & 2.22-13 \\
 $3d^8~  ^3F^e 4p$ & $^2D^o$ & 2.85-13& $3d^8~  ^3P^e 4p$ & $^2D^o$ & 2.88-13& $
3d^74s ^3F^e 4p$ & $^2F^o$ & 8.63-14& $3d^74s ^3F^e 4p$ & $^2G^o$ & 1.26-13 \\
 $3d^9           $ & $^2D^e$ & 2.43-13& $3d^8~  ^3P^e 4p$ & $^2P^o$ & 1.91-13& $
3d^8~  ^3P^e 4s$ & $^2P^e$ & 8.51-14& $3d^74s ^3F^e 4p$ & $^2D^o$ & 7.19-14 \\
 $3d^8~  ^3F^e 4p$ & $^2F^o$ & 2.33-13& $3d^8~  ^1D^e 4p$ & $^2P^o$ & 1.02-13& $
3d^8~  ^3F^e 4p$ & $^2G^o$ & 6.21-14& $3d^74s ^3F^e 4p$ & $^2F^o$ & 5.53-14 \\
 $3d^8~  ^3F^e 4p$ & $^2G^o$ & 1.37-13& $3d^8~  ^1G^e 4s$ & $^2G^e$ & 1.01-13& $
 3d^74s^2      $ & $^2P^e$ & 5.66-14& $3d^8  ~^3F^e 4p$ & $^2D^o$ & 4.17-14 \\
 $3d^8~  ^1G^e 4s$ & $^2G^e$ & 1.35-13& $3d^74s^2       $ & $^2H^e$ & 9.46-14& $
3d^8  ~^3F^e 4p$ & $^2D^o$ & 5.65-14& $3d^8  ~^3F^e 4p$ & $^2G^o$ & 4.13-14 \\
 $3d^8~  ^3F^e 4s$ & $^2F^e$ & 1.35-13& $3d^8~  ^3F^e 4p$ & $^2D^o$ & 8.76-14& $
3d^8  ~^3P^e 4p$ & $^2D^o$ & 4.20-14& $3d^8  ~^3P^e 4s$ & $^2P^e$ & 3.33-14 \\
 $3d^8~  ^3P^e 4p$ & $^2P^o$ & 8.02-14& $3d^8~  ^3F^e 4p$ & $^2F^o$ & 7.51-14& $
3d^9           $ & $^2D^e$ & 4.08-14& $3d^8  ~^3F^e 4p$ & $^2F^o$ & 2.88-14 \\
 $3d^8~  ^3F^e 5p$ & $^2G^o$ & 7.78-14& $3d^9           $ & $^2D^e$ & 7.29-14& $
3d^8  ~^3F^e 4p$ & $^2F^o$ & 3.75-14& $3d^8  ~^3P^e 4p$ & $^2D^o$ & 2.26-14 \\
 $3d^8~  ^3P^e 4p$ & $^2D^o$ & 7.50-14& $3d^8~  ^3P^e 5p$ & $^2D^o$ & 5.24-14& $
3d^8  ~^3P^e 4p$ & $^2P^o$ & 2.38-14& $3d^74s^2       $ & $^2P^e$ & 2.19-14 \\
 $3d^74s ^3F^e 4p$ & $^2G^o$ & 6.84-14& $3d^8~  ^3F^e 4p$ & $^2G^o$ & 4.43-14& $
3d^8  ~^1D^e 4p$ & $^2P^o$ & 1.66-14& $3d^74s ^3G^e 4p$ & $^2G^o$ & 1.94-14 \\
 $3d^74s^2       $ & $^2D^e$ & 6.21-14& $3d^8~  ^3F^e 4s$ & $^2F^e$ & 4.35-14& $
3d^8  ~^1G^e 4p$ & $^2G^o$ & 1.51-14& $3d^8  ~^3F^e 4s$ & $^2F^e$ & 1.74-14 \\
 $3d^8~  ^3F^e 5p$ & $^2D^o$ & 6.20-14& $3d^74s ^3P^e 4p$ & $^2P^o$ & 3.97-14& $
3d^8  ~^1G^e 4s$ & $^2G^e$ & 1.49-14& $3d^8  ~^3P^e 4p$ & $^2P^o$ & 1.18-14 \\
 $3d^8~  ^3F^e 5p$ & $^2F^o$ & 5.62-14& $3d^74s ^3F^e 4p$ & $^2G^o$ & 3.84-14& $
3d^8  ~^3F^e 4d$ & $^2G^e$ & 1.32-14& $3d^8  ~^1G^e 4p$ & $^2F^o$ & 8.99-15 \\
 $3d^8~  ^1D^e 4p$ & $^2P^o$ & 5.19-14& $3d^8~  ^1G^e 4p$ & $^2F^o$ & 3.40-14& $
3d^8  ~^3F^e 4s$ & $^2F^e$ & 1.19-14& $3d^8  ~^1G^e 4p$ & $^2G^o$ & 8.88-15 \\
 $3d^8~  ^3F^e 5f$ & $^2I^o$ & 4.44-14& $3d^8~  ^1D^e 4d$ & $^2P^e$ & 2.70-14& $
3d^8  ~^1G^e 4p$ & $^2F^o$ & 1.11-14& $3d^8  ~^1D^e 4p$ & $^2P^o$ & 8.39-15 \\
 $3d^8~  ^3F^e 4d$ & $^2H^e$ & 4.35-14& $3d^8~  ^1D^e 5d$ & $^2P^e$ & 2.69-14& $
3d^8  ~^1D^e 4p$ & $^2D^o$ & 1.05-14& $3d^8  ~^3F^e 4f$ & $^2I^o$ & 8.32-15 \\
 $3d^74s4p       $ & $^2H^o$ & 4.17-14& $3d^8~  ^3F^e 5p$ & $^2G^o$ & 2.40-14& $
 3d^74s^2      $ & $^2H^e$ & 9.35-15& $3d^74s^2       $ & $^2F^e$ & 8.03-15 \\
 $3d^8~  ^3F^e 5d$ & $^2H^e$ & 4.02-14& $3d^74s^2       $ & $^2D^e$ & 2.17-14& $
3d^74s ^3G^e 4p$ & $^2G^o$ & 8.30-15& $3d^8~  ^1G^e 5p$ & $^2G^o$ & 7.82-15 \\
 Sum= & & 2.67-12 & & & 2.47-12 & & & 9.15-13 & & & 1.25-12 \\
 Total= & & 5.29-11 & & & 1.85-11 & & & 4.28-12 & & & 3.69-12 \\
 \multicolumn{2}{l}{\%contribution=} &  5\% & & & 13\% & & & 21\% & & & 34\% \\
\multicolumn{12}{c}{Quartets} \\
 $3d^74s ^5F^e 4p$ & $^4G^o$ & 1.20-11& $3d^74s ^5F^e 4p$ & $^4G^o$ & 4.38-12& $
3d^74s ^5F^e 4p$ & $^4G^o$ & 4.80-13& $3d^74s ^5F^e 4p$ & $^4G^o$ & 2.09-13 \\
 $3d^74s ^5F^e 4p$ & $^4F^o$ & 6.68-12& $3d^74s ^5F^e 4p$ & $^4F^o$ & 2.69-12& $
3d^74s ^5F^e 4p$ & $^4F^o$ & 4.16-13& $3d^74s ^5F^e 4p$ & $^4F^o$ & 1.95-13 \\
 $3d^8~  ^3F^e 4p$ & $^4G^o$ & 3.36-12& $3d^8~  ^3F^e 4p$ & $^4G^o$ & 1.46-12& $
3d^74s ^5F^e 4p$ & $^4D^o$ & 2.73-13& $3d^74s ^5F^e 4p$ & $^4D^o$ & 1.35-13 \\
 $3d^8~  ^3F^e 4p$ & $^4F^o$ & 1.41-12& $3d^8~  ^3F^e 4p$ & $^4F^o$ & 7.75-13& $
3d^8  ~^3F^e 4p$ & $^4D^o$ & 2.05-13& $3d^8  ~^3F^e 4p$ & $^4G^o$ & 1.32-13 \\
 $3d^8~  ^3F^e 4p$ & $^4D^o$ & 7.07-13& $3d^8~  ^3F^e 4p$ & $^4D^o$ & 6.93-13& $
3d^8  ~^3F^e 4p$ & $^4G^o$ & 1.95-13& $3d^74s ^3F^e 4p$ & $^4F^o$ & 1.32-13 \\
 $3d^74s ^5F^e 4p$ & $^4D^o$ & 4.08-13& $3d^74s ^5F^e 4p$ & $^4D^o$ & 4.56-13& $
3d^8  ~^3F^e 4p$ & $^4F^o$ & 1.31-13& $3d^8  ~^3F^e 4p$ & $^4D^o$ & 1.05-13 \\
 $3d^8~  ^3F^e 5p$ & $^4G^o$ & 1.49-13& $3d^8~  ^3F^e 5p$ & $^4F^o$ & 6.44-14& $
3d^8  ~^3P^e 4p$ & $^4D^o$ & 8.47-14& $3d^8  ~^3F^e 4p$ & $^4F^o$ & 9.90-14 \\
 $3d^8~  ^3F^e 4d$ & $^4H^e$ & 1.02-13& $3d^8~  ^3F^e 5p$ & $^4G^o$ & 6.36-14& $
3d^74s ^5F^e 4f$ & $^4D^o$ & 8.07-14& $3d^8  ~^3F^e 4s$ & $^4F^e$ & 7.81-14 \\
\tableline
 \end{tabular}
 \label{table3}
 \end{table}

\begin{table}
\noindent{Table 3 continues. \\ }
\scriptsize
\begin{tabular}{lcclcclcclcc}
\tableline
\multicolumn{3}{c}{100\,K} & \multicolumn{3}{c}{1000\,K}&
\multicolumn{3}{c}{10000\,K} & \multicolumn{3}{c}{20000\,K} \\
\multicolumn{2}{c}{State} & $\alpha_R$ & \multicolumn{2}{c}{State} &
$\alpha_R$ & \multicolumn{2}{c}{State} & $\alpha_R$ &
\multicolumn{2}{c}{State} & $\alpha_R$ \\
\tableline
 $3d^8~  ^3F^e 4f$ & $^4I^o$ & 1.00-13& $3d^8~  ^3P^e 4p$ & $^4P^o$ & 6.02-14& $
3d^74s ^3G^e 4p$ & $^4F^o$ & 5.55-14& $3d^74s ^3G^e 4p$ & $^4F^o$ & 6.92-14 \\
 $3d^8~  ^3F^e 5p$ & $^4F^o$ & 9.01-14& $3d^8~  ^3P^e 4p$ & $^4D^o$ & 5.70-14& $
3d^8  ~^3F^e 4s$ & $^4F^e$ & 4.61-14& $3d^74s ^3F^e 4p$ & $^4G^o$ & 6.15-14 \\
 $3d^8~  ^3F^e 5f$ & $^4I^o$ & 8.82-14& $3d^8~  ^3F^e 4d$ & $^4H^e$ & 3.12-14& $
3d^74s ^3F^e 4p$ & $^4F^o$ & 3.99-14& $3d^74s ^3G^e 4p$ & $^4G^o$ & 5.51-14 \\
 $3d^8~  ^3F^e 5d$ & $^4H^e$ & 8.81-14& $3d^8~  ^3F^e 4f$ & $^4I^o$ & 3.05-14& $
3d^74s ^3G^e 4p$ & $^4G^o$ & 3.82-14& $3d^74s ^3F^e 4p$ & $^4F^o$ & 5.07-14 \\
 $3d^74s ^3H^e 4p$ & $^4H^o$ & 8.53-14& $3d^8~  ^3F^e 6p$ & $^4G^o$ & 2.72-14& $
3d^74s ^3F^e 4p$ & $^4F^o$ & 3.65-14& $3d^74s ^5F^e 4f$ & $^4D^o$ & 4.72-14 \\
 $3d^8~  ^3F^e 4d$ & $^4G^e$ & 7.90-14& $3d^8~  ^3F^e 5d$ & $^4H^e$ & 2.69-14& $
3d^74s ^5P^e 4p$ & $^4D^o$ & 3.08-14& $3d^8  ~^3P^e 4p$ & $^4D^o$ & 4.53-14 \\
 $3d^8~  ^3F^e 5f$ & $^4H^o$ & 7.67-14& $3d^8~  ^3F^e 5f$ & $^4I^o$ & 2.69-14& $
3d^74s ^3F^e 4p$ & $^4G^o$ & 2.90-14& $3d^8  ~^3F^e 4f$ & $^4H^o$ & 4.26-14 \\
 $3d^8~  ^3F^e 6p$ & $^4G^o$ & 7.28-14& $3d^8~  ^3F^e 5p$ & $^4D^o$ & 2.64-14& $
3d^8  ~^3P^e 4p$ & $^4P^o$ & 2.72-14& $3d^74s ^5P^e 4p$ & $^4D^o$ & 3.60-14 \\
 $3d^8~  ^3F^e 5p$ & $^4D^o$ & 7.21-14& $3d^74s ^3H^e 4p$ & $^4H^o$ & 2.58-14& $
3d^74s ^3F^e 4p$ & $^4D^o$ & 2.40-14& $3d^74s ^3G^e 4p$ & $^4H^o$ & 3.59-14 \\
 $3d^8~  ^3F^e 5d$ & $^4G^e$ & 7.05-14& $3d^8~  ^3F^e 4d$ & $^4G^e$ & 2.41-14& $
3d^74s ^3G^e 4p$ & $^4H^o$ & 2.40-14& $3d^74s ^3F^e 4p$ & $^4G^o$ & 3.41-14 \\
 $3d^8~  ^3F^e 4f$ & $^4G^o$ & 6.68-14& $3d^8~  ^3F^e 5f$ & $^4H^o$ & 2.32-14& $
3d^8  ~^3F^e 5p$ & $^4F^o$ & 1.99-14& $3d^74s ^3F^e 4p$ & $^4D^o$ & 3.18-14 \\
 $3d^74s^2       $ & $^4F^e$ & 6.66-14& $3d^8~  ^3F^e 5d$ & $^4G^e$ & 2.16-14& $
3d^74s ^3F^e 4p$ & $^4G^o$ & 1.95-14& $3d^74s ^3H^e 4p$ & $^4I^o$ & 2.43-14 \\
 Sum= & & 2.58-11 & & & 1.10-11 & & & 2.26-12 & & & 1.62-12 \\
 Total= & & 5.29-11 & & & 1.85-11 & & & 4.28-12 & & & 3.69-12 \\
 \multicolumn{2}{l}{\%contribution=} & 49\% & & & 59\% & & & 53\% & & & 44\% \\
\tableline
 \end{tabular}
 \label{table3}
 \end{table}

\begin{table}
\caption{Total recombination rate coefficients, $\alpha_R(T)$, in units
of $cm^3s^{-1}$, for e + Ni III $\rightarrow$ Ni II. The temperature is 
given in K.
\label{table4}}
\begin{tabular}{cccccc}
\tableline
$log_{10}T$ & $\alpha_R$ & $log_{10}T$ & $\alpha_R$ & $log_{10}T$ & $\alpha_R$
\\
\tableline
 1.0 &  1.78E-10 & 3.1 &   1.61E-11 & 5.2 &  4.15E-12 \\
 1.1 &  1.57E-10 & 3.2 &   1.40E-11 & 5.3 &  3.49E-12 \\
 1.2 &  1.38E-10 & 3.3 &   1.21E-11 & 5.4 &  2.85E-12 \\
 1.3 &  1.21E-10 & 3.4 &   1.04E-11 & 5.5 &  2.27E-12 \\
 1.4 &  1.07E-10 & 3.5 &   8.93E-12 & 5.6 &  1.77E-12 \\
 1.5 &  9.45E-11 & 3.6 &   7.66E-12 & 5.7 &  1.36E-12 \\
 1.6 &  8.36E-11 & 3.7 &   6.57E-12 & 5.8 &  1.03E-12 \\
 1.7 &  7.41E-11 & 3.8 &   5.64E-12 & 5.9 &  7.68E-13 \\
 1.8 &  6.59E-11 & 3.9 &   4.87E-12 & 6.0 &  5.69E-13 \\
 1.9 &  5.90E-11 & 4.0 &   4.28E-12 & 6.1 &  4.19E-13 \\
 2.0 &  5.29E-11 & 4.1 &   3.86E-12 & 6.2 &  3.08E-13 \\
 2.1 &  4.78E-11 & 4.2 &   3.66E-12 & 6.3 &  2.19E-13 \\
 2.2 &  4.34E-11 & 4.3 &   3.69E-12 & 6.4 &  1.59E-13 \\
 2.3 &  3.96E-11 & 4.4 &   3.94E-12 & 6.5 &  1.15E-13 \\
 2.4 &  3.61E-11 & 4.5 &   4.35E-12 & 6.6 &  8.28E-14 \\
 2.5 &  3.28E-11 & 4.6 &   4.82E-12 & 6.7 &  5.97E-14 \\
 2.6 &  2.97E-11 & 4.7 &   5.23E-12 & 6.8 &  4.30E-14 \\
 2.7 &  2.67E-11 & 4.8 &   5.47E-12 & 6.9 &  3.09E-14 \\
 2.8 &  2.38E-11 & 4.9 &   5.47E-12 & 7.0 &  2.22E-14 \\
 2.9 &  2.11E-11 & 5.0 &   5.21E-12 &    &           \\
 3.0 &  1.85E-11 & 5.1 &   4.75E-12 &    &           \\
\tableline
\end{tabular}
\end{table}

\newpage

\begin{figure}
\centering
\psfig{figure=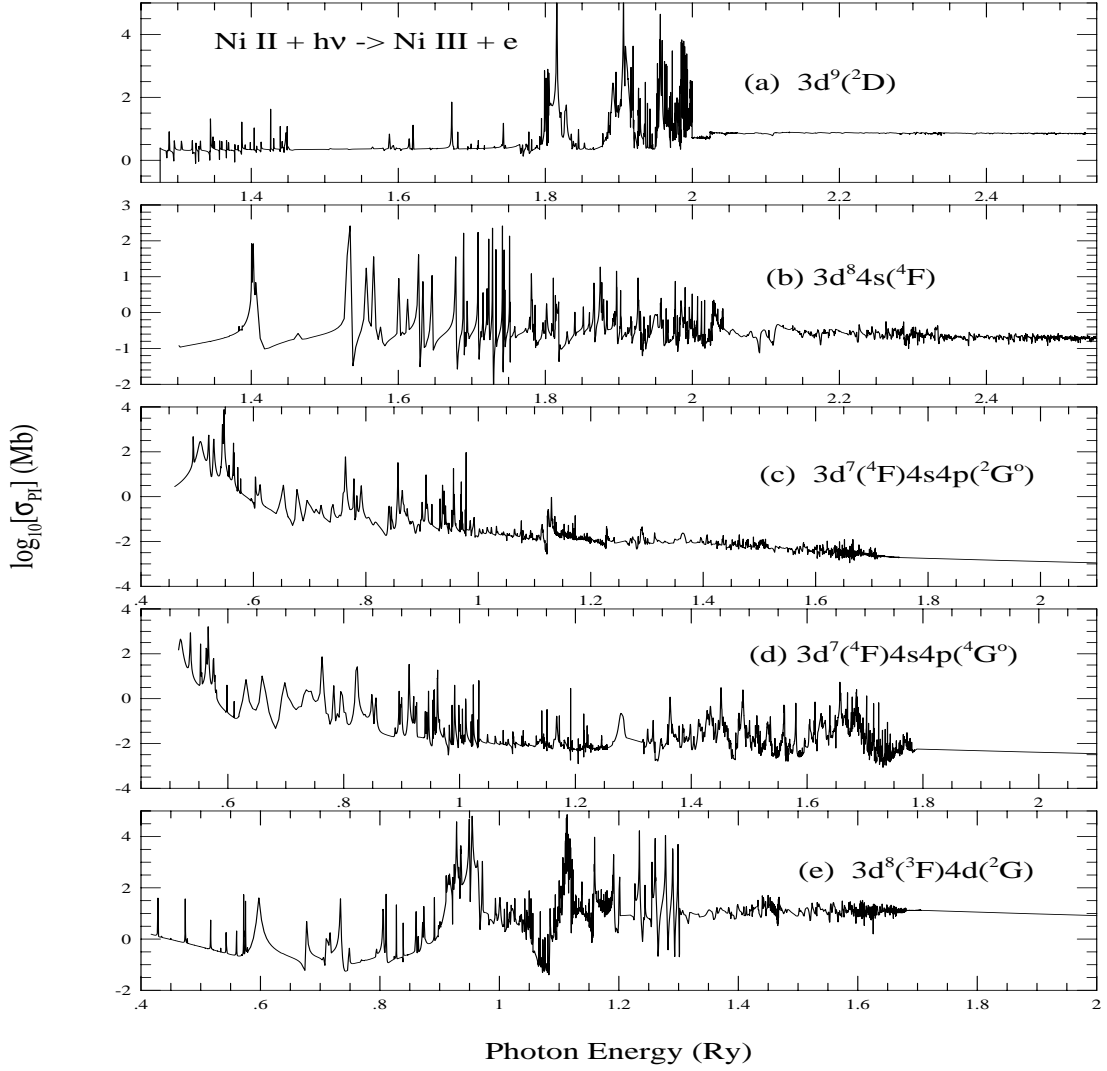,height=17.0cm,width=18.0cm}
\caption{Partial photoionization cross sections, $\sigma_{PI}$, of the 
ground state, (a) $3d^9(^2D)$, and excited states, (b) $3d^84s(^4F)$, (c) 
$3d^7(^4F)4s4p(^2G^o)$, (d) $3d^7(^4F)4s4p(^4G^o)$, and (e)
$3d^8(^3F)4d(^2G)$ of Ni II leaving the core in the ground state 
$3d^8(^3F)$. }
\end{figure}

\begin{figure}
\centering
\psfig{figure=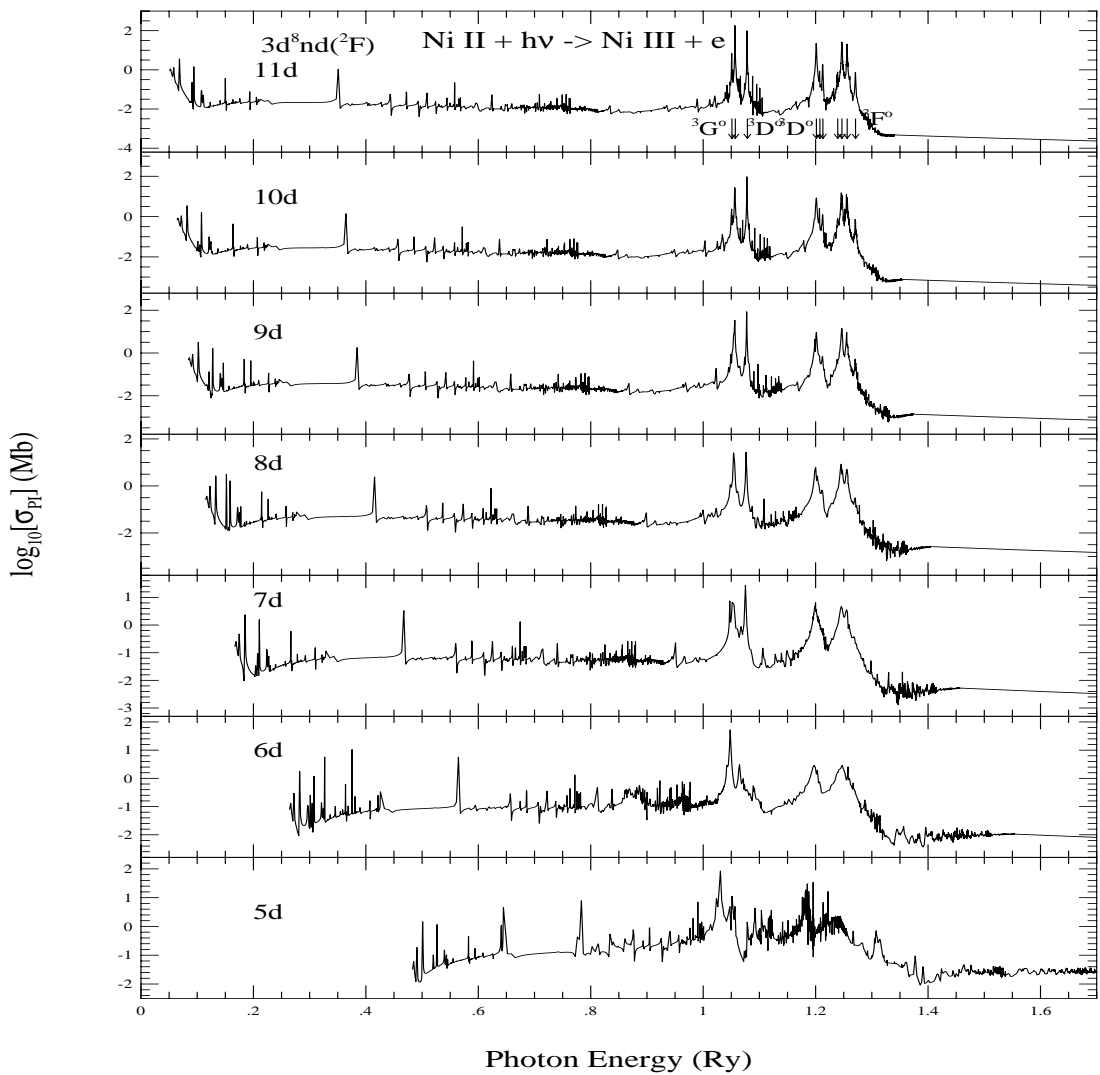,height=17.0cm,width=18.0cm}
\caption{Partial photoionization cross sections, $\sigma_{PI}$, of the 
Rydberg series of states, $3d^8nd(^2F)$, $5d \leq nd \leq 11d$,
of Ni II illustrating PEC resonances at energies pointed by arrows in 
the top panel. PECs are manifested by dipole allowed transition in the
core from the core ground state. }
\end{figure}

\begin{figure}
\centering
\psfig{figure=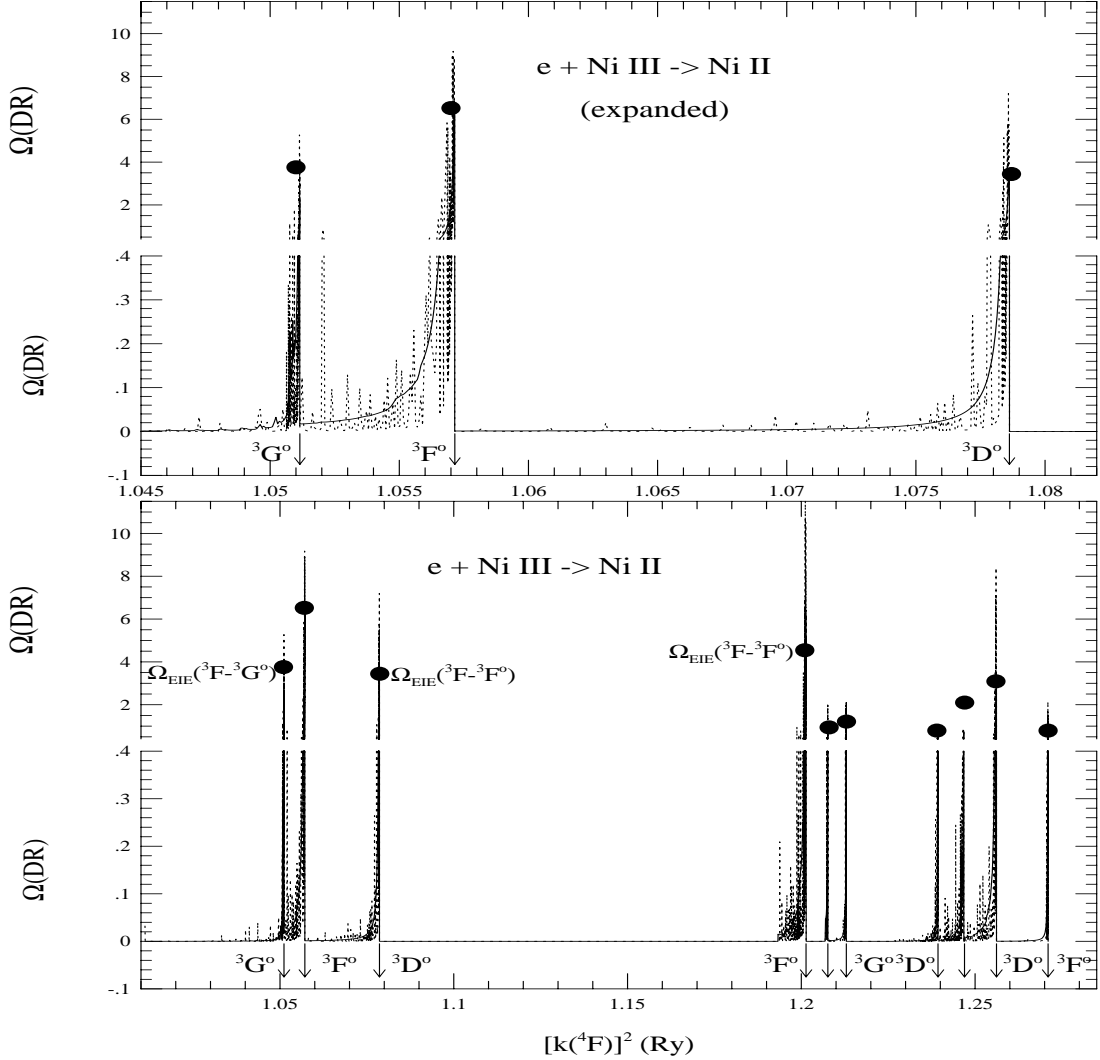,height=17.0cm,width=18.0cm}
\caption{DR collision strength, $\Omega(DR)$, for recombination of 
e + Ni III $\rightarrow$ Ni II: detailed with resonances (dotted 
curves) and resonance averaged (solid curves). The arrows point the
energy positions of the 10 target thresholds $3d^74p(z^3G^o$, $z^3F^o$,
$z^3D^o$, $y^3F^o$, $y^3D^o$, $y^3G^o$, $x^3D^o$, $x^3G^o$, $w^3D^o$,
$x^3F^o)$ where DR resonances are converging in the bottom panel while
the top panel presents an expanded detail of the first 3 core thresholds. 
The filled circles are the excitation collision strength, $\Omega(EIE)$, 
at these thresholds.
 }
\end{figure}

\begin{figure}
\centering
\psfig{figure=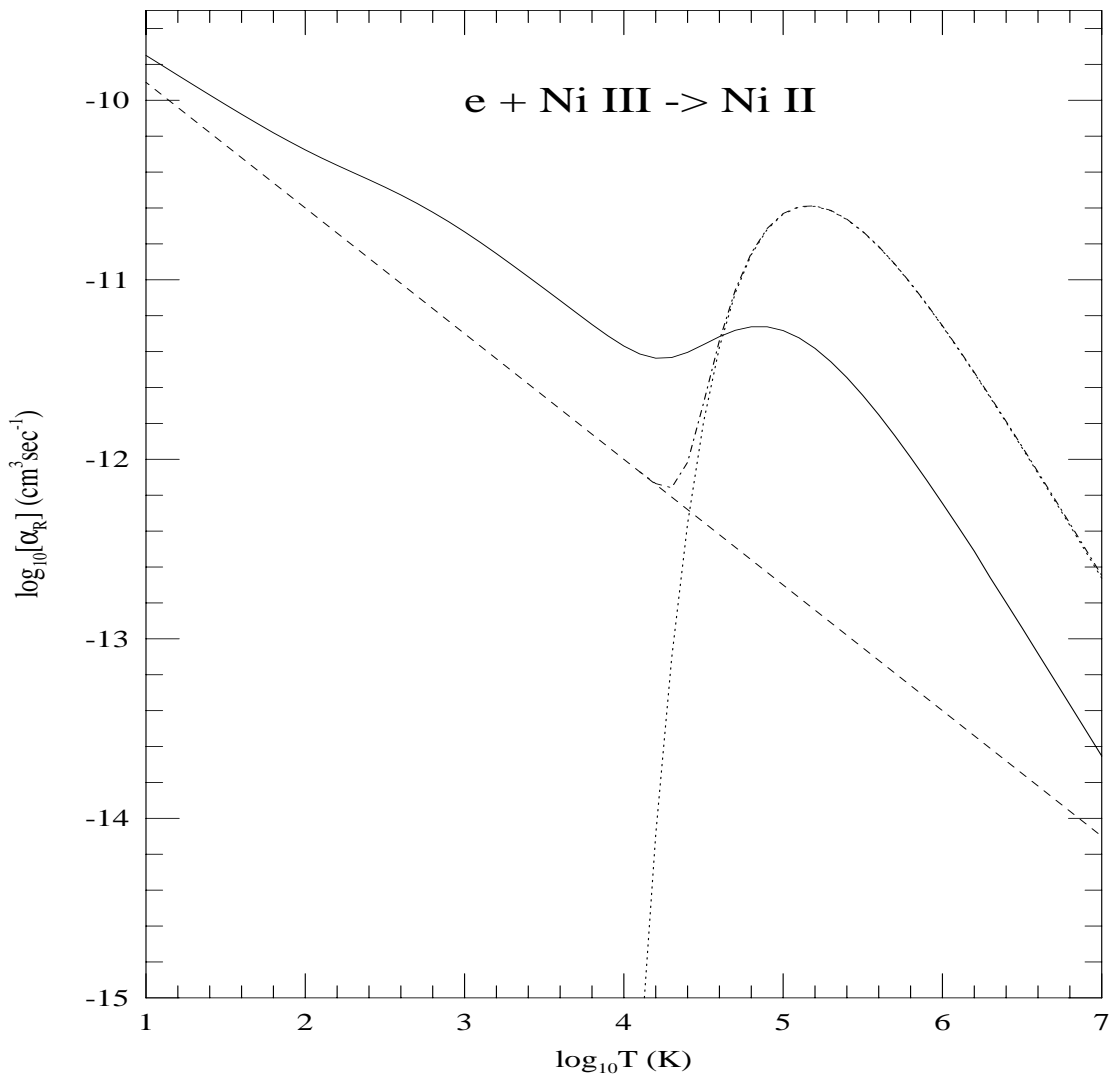,height=17.0cm,width=18.0cm}
\caption{Total electron-ion recombination rate coefficients, 
$\alpha_R(T)$, for e + Ni III $\rightarrow$ Ni II of the present work
(solid curve). The dashed curve presents the RR, dashed curve the DR,
and dot-dashed curve the total rates by Shull and Steenberg (1982). 
Low-T recombination is enhanced due to near threshold resonances, but
the high-T rate is considerably reduced primarily due to autoionization
at the large number of excited states. }
\end{figure}

\end{document}